\documentclass[letterpaper,english,unsortedaddress]{revtex4}
\usepackage[T1]{fontenc}
\usepackage[latin1]{inputenc}
\usepackage{graphicx}
\usepackage{amssymb}

\makeatletter

\providecommand{\boldsymbol}[1]{\mbox{\boldmath $#1$}}

\providecommand{\tabularnewline}{\\}

\usepackage{bm}

\usepackage{babel}
\makeatother
\begin{document}

\title{Deterministic Derivation of NonEquilibrium Free Energy Theorems for
Natural Isothermal Isobaric Systems}

\author{Stephen R. Williams}

\email{swilliams@rsc.anu.edu.au}

\affiliation{Research School of Chemistry, The Australian National University,
Canberra, ACT 0200, Australia}

\author{Debra J. Searles}

\email{d.bernhardt@griffith.edu.au}

\affiliation{Nanoscale Science and Technology Centre, School of Science, Griffith
University, Brisbane, Qld 4111, Australia}

\author{Denis J. Evans}

\email{evans@rsc.anu.edu.au}

\affiliation{Research School of Chemistry, The Australian National University,
Canberra, ACT 0200, Australia}

\begin{abstract}
The nonequilibrium free energy theorems show how distributions of
work along nonequilibrium paths are related to free energy differences
between the equilibrium states at the end points of these paths. In
this paper we develop a natural way of barostatting a system and give
the first deterministic derivation of the Crooks and Jarzynski relations
for these isothermal isobaric systems. We illustrate these relations
by applying them to molecular dynamics simulations of a model polymer
undergoing stretching.
\end{abstract}
\maketitle

\section{Introduction}

Recently a set of revolutionary statistical mechanical theorems have
been derived. These theorems involve distributions of work and dissipation
along thermodynamically nonequilibrium paths. The theorems are remarkable
in that they constitute some of the very few exact statistical mechanical
relations for systems that are possibly far from equilibrium. Close
to equilibrium one of these relations (the Evans Searles Fluctuation
Theorem) can be used to derive Green-Kubo relations for linear transport
coefficients. A second remarkable feature of these theorems is that
they enable us to derive thermodynamic statements about small systems.
The thermodynamic limit is not required. The third remarkable feature
of some of the theorems is that one can derive exact relations about
equilibrium free energy differences by analysing nonequilibrium thermodynamic
path integrals.

The first relationship was presented in 1993 by Evans et al. \cite{Evans-PRL-93}
where an expression concerning the probability distribution of values
of the dissipation in the steady state was given, and tested. This
work motivated a number of papers in which various fluctuation theorems
were derived, the first of which were the Evans-Searles Transient
Fluctuation Theorem \cite{Evans-PRE-94}, and the Gallavotti-Cohen
Fluctuation Theorem \cite{Gallavotti-PRL-95}. Later the Jarzynski
Equality (JE) \cite{Jarzynski-PRL-97} (also known as the work relation
or nonequilibrium work relation) and the Crooks Fluctuation Theorem
(CFT) \cite{Crooks-JSP-98} (also known as the Crooks Identity or
Crooks Fluctuation Relation) were developed, giving expressions for
distributions of time-integrated work along nonequilibrium paths that
dynamically connect two different equilibrium thermodynamic states,
and the free energy difference between the states. In the present
paper we shall be concerned with the last two such theorems. 

There are proofs in the literature for the Jarzynski Equality in the
isothermal isobaric ensemble for stochastic dynamics \cite{Park-JCP-04}
and for deterministic dynamics thermostatted homogeneously with a
synthetic thermostat \cite{Cuendet-JCP-06}. We present the \textbf{}first
derivation \textbf{}of the Crooks and Jarzynski relations for time
reversible deterministic, systems \textbf{}which have a state point
controlled by an external \textbf{}constant temperature, constant
pressure reservoir. In order to carry out this derivation we surround
a natural system which obeys Newtonian mechanics, the so-called {}``system
of interest'', with a reservoir region of fixed temperature and which
maintains a constant pressure on the system of interest. \textbf{}This
is a new model, and for many experimental systems, (i.e. those held
at fixed pressure and temperature) it provides the most realistic
arrangement yet available for simulating a process occurring under
these conditions because it separates the reservoir from the system
of interest.

The existing statistical mechanical proofs \cite{Evans-molphys-03,Cuendet-PRL-06,scholl-JCP-06,Cuendet-JCP-06}
of the JE \cite{Jarzynski-PRL-97,Jarzynski-PRE-97} are for the change
in Helmholtz free energy for a system held at constant volume and
in contact with a thermostat. When applied to an experiment, held
at constant volume, the pressure of the initial equilibrium phase
may be significantly different from the final equilibrium phase. Frequently
experiments are performed at constant pressure rather than constant
volume. Therefore it is often more appropriate to use the isothermal
isobaric ensemble and the Gibbs free energy \cite{Liphardt-sci-02,Collin-Nat-05}.
Stochastic derivations, based on the Markovian assumption, exist for
this case \cite{Jarzynski-PRE-97,Crooks-JSP-98,Hummer-pnas-01,Park-JCP-04}.
However, most actual experimental systems do not satisfy Markovian
assumptions \textbf{}as the basic equations of motion are not stochastic
\textbf{}\cite{zwanzig-book,Evans-Morriss-book}. Stochastic derivations
based on Markovian assumptions when applied to deterministic systems,
are only valid for systems that obey the Langevin equation with white
noise only \cite{Hansen-book}. \textbf{}Hummer and Szabo \cite{Hummer-pnas-01,Hummer-acr-05}
present a formal derivation using Feynman-Kac theory and point out
that their formalism remains valid if Hamiltonian or Schr\"odinger
operators are used. At equilibrium such operators do not generate
an ergodic canonical distribution because states with different energies
never mix. When work is performed on the system they do not provide
a mechanism for the dissipation of heat, and therefore although they
are a useful theoretical construct for obtaining the JE or CFT, the
nonequilibrium dynamics is not representative of a system, on which
work is performed, which is in contact with a large thermal reservoir.
The invocation of the initial canonical distribution implicitly assumes
contact with a thermal reservoir. The use of thermostats provide a
mechanism by which the JE and CFT can be rigorously obtained for deterministic
systems, and provides a realistic model of natural experimental systems
where the system typically relaxes to the same temperature and pressure
as it was at initially. \textbf{}Following the approach in \cite{Evans-molphys-03}
we will derive the CFT \cite{Crooks-PRE-99} and then the JE, which
follows trivially. 

In the system we consider, the reservoir region obeys time reversible
Newtonian equations of motion that are augmented with unnatural thermostatting
and barostatting terms, while the system of interest obeys the natural
time reversible Newtonian equations. We argue that as the number of
degrees of freedom in the reservoir becomes much larger than that
of the system of interest, and as the physical separation of these
reservoirs from the system of interest becomes larger the system of
interest looses knowledge of the details of how the thermostatting
and barostatting is achieved \cite{Williams-PRE-04}. The system of
interest simply {}``sees'' that it is surrounded by a large time
reversible reservoir region that is in thermodynamic equilibrium at
a known temperature and pressure. The resulting Crooks and Jarzynski
relations refer only to the work done on the system of interest and
only the temperature and pressure of the reservoir. Hence we expect
that the derived results of this gedanken experiment will apply to
naturally occurring systems. While thermostats which only act on particles
which form a solid wall enclosing the fluid are widely used, (for
example see \cite{Ayton-JCP-01,Evans-adv-phys-02,Williams-PRE-04}),
the development of a natural barostat is new.

Here we develop equations of motion for a fluid maintained at an externally
controlled temperature and pressure. In contrast with previous work
on wall thermostatted systems \cite{Ayton-JCP-01,Evans-adv-phys-02,Williams-PRE-04},
the particles are not identified as {}`wall particles' or {}`fluid
particles', but the system and reservoir (which could be fluid and
wall) are distinguished by their location. In fact, although the scheme
can be applied to systems within thermostatting walls as before, it
does not require any solid walls. As in previous work, the system
of interest is purely Hamiltonian. Thus we arrive at the possibility
of a fluid, large enough to contain a system of interest, which is,
by the physical principle of locality, identical to a fluid regulated
by a large external heat reservoir. However for this newly constructed
fluid the effective decoupling from the thermal reservoir is axiomatic.
This allows all of the previously developed work, on thermostatted
dynamics, to be applied directly to this system. Thus given the principle
of locality we obtain rigorous derivations for linear and nonlinear
response theory, the fluctuation theorem, Le Chatelier's principle,
Green Kubo relations, the CFT, the JE and the second law of thermodynamics
\cite{Evans-Morriss-book,Evans-adv-phys-02}.

\section{\textup{\normalsize The Equations of Motion}}

\begin{figure}
\resizebox{8.5cm}{!}{\includegraphics{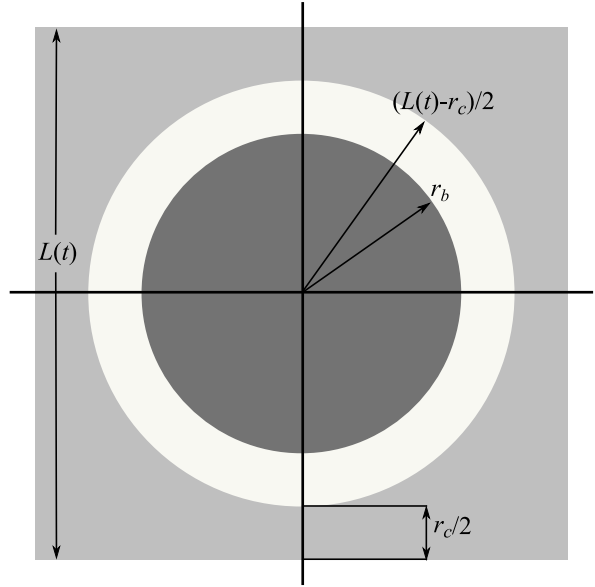}}

\caption{The diagram represents the three regions in the simulation. The inner
darker circle represents the Hamiltonian region which is not affected
by the barostat or the thermostat and whose diameter is held fixed,
the white region is where the function $g(\boldsymbol{q},V)$ increases
from zero to unity and the outer lighter grey region is where $g(\boldsymbol{q},V)$
is set to unity. The minimum width separating the white region from
it's minimum image must be at least as wide as the maximum cutoff
radius for the interaction between the particles. The outer boundary
of the white region may fluctuate with the volume. \label{figure-Sim-Cell}}
\end{figure}

We wish to set up a dynamical model of a molecular system in the isobaric
isothermal ensemble. The system of interest, where the equations of
motion are Newtonian, will be surrounded by a reservoir region that
uses Nos\'e Hoover feedback \cite{Nose-JCP-84,Nose-mp-84,Hoover-PRA-85}
to control the pressure and the temperature. This feedback mechanism
is unnatural, but deterministic and time reversible. The number of
particles in the full system (reservoir plus system of interest),
$N$, is constant. This can be achieved by making the total system
periodic or by surrounding it by boundaries that prevent escape of
particles. In the results presented below, we use periodic boundary
conditions with a cubic unit cell of length $L(t)$ centred as shown
in Figure \ref{figure-Sim-Cell}. We fix the total momentum of the
full system to zero ($\mathbf{p}_{tot}=\sum_{i=1}^{N}\mathbf{p}_{i}=\mathbf{0}$)
and the unnormalized centre of mass $\mathbf{R}$ of the full system
to a fixed position which streams with the cell dilation, $\mathbf{R}(t)=\sum_{i=1}^{N}m_{i}\mathbf{q}_{i}(t)=\mathbf{R}(0)\, L(t)/L(0)$,
\textbf{}by using appropriate equations of motion. Obviously if the
centre of mass is set at the origin it will remain there, $\boldsymbol{R}(t)=0\:\:\forall\:\: t$,
and for convenience we will restrict ourselves to this case. In order
to specify the thermostatting and natural regions, we define $q_{i}$
as the magnitude of the displacement of particle $i$ from the centre
of its unit cell which in the case of the central cell is fixed at
the origin $\mathbf{q}_{0}=0$ \cite{foot1}. In mathematical form
$q_{i}=|(\mathbf{q}_{i}-\mathbf{q}_{0})_{mod\: L}|$. When the distance
of a particle from the centre of its unit cell, $q_{i}$, is less
than $r_{b}$, the dynamics are Newtonian (i.e. natural), see Figure
\ref{figure-Sim-Cell}. 

The equations of motion are\begin{eqnarray}
\dot{\mathbf{q}}_{i} & = & \mathbf{p}_{i}/m+\alpha_{V}g(\mathbf{q}_{i},V)\mathbf{q}_{i}-\boldsymbol{\gamma}_{q}\nonumber \\
\dot{\mathbf{p}}_{i} & = & \mathbf{F}_{i}-\alpha_{V}g(\mathbf{q}_{i},V)\mathbf{p}_{i}-\alpha_{T}g(\mathbf{q}_{i},V)\mathbf{p}_{i}-\boldsymbol{\gamma}_{p}\label{EOM}\\
\dot{V} & = & D\alpha_{V}V.\nonumber \end{eqnarray}
where $D$ is the Cartesian dimension. For a unit cell that is cubic,
the time dependence of the cell length $L=\sqrt[D]{V}$ is given by
$\dot{L}=\alpha_{V}L$. The volume and the temperature are controlled
through the multipliers $\alpha_{V}\,\&\,\alpha_{T}$. These multipliers
only operate if a particle is in the region $q_{i}>r_{b}$. The function
$g(\mathbf{q}_{i},V)$ is a position sensitive switch that smoothly
turns the thermostat and barostat off in the system of interest and
on in the reservoir region:\begin{eqnarray}
g(\mathbf{q}_{i},V(t)) & = & \frac{1}{2}\left(1-\cos\left(\pi\frac{q_{i}-r_{b}}{(L-r_{c})/2-r_{b}}\right)\right),\;\; r_{b}<q_{i}<(L-r_{c})/2\nonumber \\
g(\mathbf{q}_{i},V(t)) & = & 1,\;\; q_{i}>(L-r_{c})/2\label{f(q)}\\
g(\mathbf{q}_{i},V(t)) & = & 0,\;\; q_{i}<r_{b}.\nonumber \end{eqnarray}

In this equation $r_{c}$ is some distance equal to or greater than
the longest distance over which the atoms interact. This means any
work done on the system due to the volume $V$ changing is accounted
for by the equations of motion. Of course our choice for $g(\mathbf{q}_{i},V(t))$
is not unique. The total momentum is constrained as stated above by
setting $\boldsymbol{\gamma}_{p}=\frac{1}{N}\sum_{j=1}^{N}(\boldsymbol{F}_{j}-(\alpha_{V}+\alpha_{T})g(\mathbf{q}_{j},V)\mathbf{p}_{j})$,
and is set to zero, and the system's centre of mass is constrained
by setting $\boldsymbol{\gamma}_{q}=\alpha_{V}\sum_{j=1}^{N}(g(\boldsymbol{q}_{j},V)-1)m_{j}\boldsymbol{q}_{j}/\sum_{j=1}^{N}m_{j}$,
and for convenience set to zero, $\boldsymbol{R}(t)=0\:\:\forall\:\: t$.
This gives Newtonian equations including the streaming produced by
dilation of the box when $g(\boldsymbol{q}_{i},V)=0$ (i.e. in the
central region), and is consistent with the usual isobaric-isothermal
equations of motion when $g(\boldsymbol{q}_{i},V)=1\:\forall\:\;(\boldsymbol{q}_{i},V)$
\cite{Evans-Morriss-book,rapaport-book}. %

The time dependencies of the multipliers are such that the equilibrium
equations of motion preserve the isothermal isobaric distribution
function with fixed centre of mass at the origin and zero total momentum:\begin{equation}
f_{0}(\boldsymbol{\Gamma},V)=\frac{\exp(-\beta(H_{0}(\boldsymbol{\Gamma})+P_{0}V)\delta(\mathbf{p}_{tot})\delta(\mathbf{R})}{\int_{0}^{\infty}dV\,\int d\boldsymbol{\Gamma}\exp(-\beta(H_{0}(\boldsymbol{\Gamma})+P_{0}V))\delta(\mathbf{p}_{tot})\delta(\mathbf{R})}.\label{f_0}\end{equation}
Here $\mathbf{\Gamma\equiv}\mathbf{(q},\mathbf{p})$, $\beta=1/(k_{B}T_{0})$
where $T_{0}$ is the thermostat temperature, $H_{0}$ is the Hamiltonian
associated with the standard Newtonian equations of motion and $P_{0}$
is the externally applied pressure. We note that this approach can
be used to produce equations of motion that preserve the isothermal
isobaric distribution while only considering configurational variables
for the feedback mechanism \cite{Braga-JCP-2006}. The details of
how the distribution function is preserved are given below, resulting
in the following time dependencies for the multipliers \begin{eqnarray}
\dot{\alpha}_{T} & = & \left(\frac{\sum_{i=1}^{N}g(\mathbf{q}_{i},V)\mathbf{p}_{i}\cdot\mathbf{p}_{i}}{Dmk_{B}T_{0}}-\left[1-\frac{1}{N}\right]\sum_{i=1}^{N}g(\mathbf{q}_{i},V)\right)\frac{1}{\tau_{T}^{2}}\nonumber \\
\dot{\alpha}_{V} & = & \left(\frac{\sum_{i}^{N}g(\mathbf{q}_{i},V)(\mathbf{p}_{i}\cdot\mathbf{p}_{i}/m+\mathbf{F}_{i}\cdot\mathbf{q}_{i})}{Dk_{B}T_{0}}+\left[1-\frac{1}{N}\right]\frac{1}{D}\sum_{i=1}^{N}\nabla_{i}g(\mathbf{q}_{i},V)\cdot\mathbf{q}_{i}+1-\frac{P_{0}V}{k_{B}T_{0}}+\xi\right)\frac{1}{\tau_{V}^{2}}\label{feed-back}\end{eqnarray}
where $\tau_{T}\,\&\,\tau_{V}$ are arbitrary time constants and $\xi=\frac{1}{DNk_{B}T}\sum_{i=1}^{N}\sum_{j=1}^{N}g(\boldsymbol{\mathbf{q}_{i}},V)\boldsymbol{q_{i}\cdot}\boldsymbol{F}_{j}$
. Usually the forces all sum to zero $\sum_{j=1}^{N}\boldsymbol{F}_{j}=0$
and thus $\xi=0$, however if an external conservative potential is
present (the case of a stationary optical trap for example \cite{wang-prl-02})
then $\xi\ne0$. Note that as usual, care needs to be taken with the
use $\sum_{i}^{N}g(\mathbf{q}_{i},V)\mathbf{F}_{i}\cdot\mathbf{q}{}_{i}$
in a periodic system, and the correct expression for this case is
given in the footnote \cite{foot2}. The constraints on the system,
$\boldsymbol{\gamma_{q}}$ and $\boldsymbol{\gamma_{p}}$ may be made
arbitrarily small by increasing the number of particles in the system
or alternatively by choosing the arbitrary time constants $\tau_{T}$
and $\tau_{V}$ to be large while keeping the number of particles
in the system fixed. The phase space compression factor, i.e. the
divergence of the equations of motion, $\Lambda=\nabla_{\boldsymbol{q}}\cdot\dot{\boldsymbol{q}}+\nabla_{\boldsymbol{p}}\cdot\dot{\boldsymbol{p}}+\partial\dot{V}/\partial V+\partial\dot{\alpha}_{T}/\partial\alpha_{T}+\partial\dot{\alpha}_{V}/\partial\alpha_{V}$,
is\begin{equation}
\Lambda=-\alpha_{T}(1-1/N)D\sum_{i=1}^{N}g(\mathbf{q}_{i},V)+\alpha_{V}(1-1/N)\sum_{i=1}^{N}\nabla_{i}g(\mathbf{q}_{i},V)\cdot\mathbf{q}_{i}+\alpha_{V}D,\label{compresion-factor}\end{equation}
and the extended Hamiltonian is defined as\begin{equation}
H_{E}=H_{0}+\frac{D}{2}\alpha_{T}^{2}\tau_{T}^{2}k_{B}T_{0}+\frac{D}{2}\alpha_{V}^{2}\tau_{V}^{2}k_{B}T_{0}.\label{He}\end{equation}
The extended instantaneous enthalpy is then defined $I_{E}\equiv H_{E}+P_{0}V$.
We are now in a position to consider the time-dependence of $f_{0}(\boldsymbol{\Gamma},V)$
using the dynamics. We expect an equilibrium distribution function
that is consistent with Boltzmann's postulate of equal a priori probability
Eq. \ref{f_0}. \textbf{}A dynamical system in equilibrium has the
property that the distribution function for the ensemble is preserved
$\partial f(\mathbf{\Gamma},V,t)/\partial t=0$, i.e. it is the time
independent solution to the Liouville equation. Using the equations
of motion it can be shown that\begin{equation}
\frac{dI_{E}}{dt}=k_{B}T_{0}\Lambda\label{Eq-condition}\end{equation}
and using,

\begin{equation}
\frac{\partial f(\boldsymbol{\Gamma},V,t)}{\partial t}=-\dot{\boldsymbol{\Gamma}}\cdot\nabla_{\boldsymbol{\Gamma}}f(\boldsymbol{\Gamma},V,t)-\dot{V}\frac{\partial f(\boldsymbol{\Gamma},V,t)}{\partial V}-\Lambda f(\boldsymbol{\Gamma},V,t),\label{Liouiville}\end{equation}
it is straightforward to show that this condition is obeyed when the
extended distribution function is of the form, $f(\mathbf{\boldsymbol{\Gamma}^{\prime}},t)\propto\exp(-\beta I_{E}(\mathbf{\Gamma^{\prime}}))$
where $\mathbf{\Gamma}^{\prime}$ represents the extended phase space
vector $\mathbf{\Gamma}^{\prime}\equiv(\mathbf{\boldsymbol{\Gamma}},V,\mathbf{\alpha}_{T},\mathbf{\alpha}_{V})$.
Thus we obtain the standard isothermal isobaric distribution function
with the addition of the two extra independent Nos\'e Hoover variables,\begin{eqnarray}
f(\boldsymbol{\Gamma}^{\prime}) & = & f_{0}(\boldsymbol{\Gamma},V)\,\frac{D\tau_{T}\tau_{V}}{2\pi}\exp\left(\frac{D}{2}\left(\alpha_{T}^{2}\tau_{T}^{2}+\alpha_{V}^{2}\tau_{V}^{2}\right)\right).\label{equib-dist-function}\end{eqnarray}

Given the standard isothermal isobaric distribution function, $f_{0}(\boldsymbol{\Gamma},V)$,
Eq. \ref{f_0}, and assuming the system undergoes a uniform dilation
it is easy to prove (i.e. the virial theorem) that at equilibrium,\begin{equation}
P_{0}=\left\langle \frac{1}{DV(t)}\left(\frac{\sum_{i=1}^{N}\mathbf{p}_{i}\cdot\mathbf{p}_{i}}{m}+\sum_{i=1}^{N-1}\sum_{j=i+1}^{N}\mathbf{F}_{ij}\cdot\mathbf{q}_{ij}\right)\right\rangle \label{Pressure-mechanical}\end{equation}
and that the kinetic temperature is given by the equipartition theorem,
\begin{equation}
k_{B}T=\frac{1}{D(N-1)m}\sum_{i=1}^{N}(\mathbf{p}_{i}\mathbf{\cdot}\mathbf{p}_{i}).\label{Temperature-kinetic}\end{equation}
So we expect these equations to be obeyed if the above is implemented
in a computer simulation.

\section{Testing For Equilibrium}

It is perhaps most effective to test for equilibrium using a small
periodic system in two Cartesian dimensions. To do this we will make
$r_{b}=0$ , so all particles lie in the reservoir region at all times.
So,\begin{eqnarray}
g(\mathbf{q}_{i},V(t)) & = & \frac{1}{2}\left(1-\cos\left(2\pi\frac{q_{i}}{L-r_{c}}\right)\right),\;\;\;0<q_{i}<(L-r_{c})/2\label{f(q)-small}\\
g(\mathbf{q}_{i},V(t)) & = & 1,\;\;\; q_{i}>(L-r_{c})/2,\nonumber \end{eqnarray}
where $L=V^{1/D}$ and periodic boundary conditions are used \cite{foot2}.

We set the multipliers $\alpha_{V}=\alpha_{T}=4$, with the Cartesian
dimension $D=2$, the number of particles $N=36$ and used the WCA
potential \cite{WCA-JCP-71} with the Lennard Jones $\sigma$ and
$\epsilon$ parameters set to unity and the potential cutoff radius
coinciding with $r_{c}$ in Eq. \ref{f(q)-small}. Two simulation
sets were computed, firstly at the pressure $P_{0}=0.5$ and temperature
$T_{0}=1$, then at the pressure $P_{0}=3.5$ and temperature $T_{0}=4$.
The equations Eq. \ref{Pressure-mechanical} and Eq. \ref{Temperature-kinetic}
were used to obtain the average pressure and temperature. At the two
state points averages were calculated for $10^{5}$ time steps with
$dt=0.001$ using a $4th$ order Runge-Kutta integrator \cite{butcher-ANM-96}.
This was repeated $50$ times and the standard error was calculated.
The results are shown in Table \ref{table: small equib}. The difference
between the input parameters and the averages is consistent with the
standard error for the four variables. This provides strong evidence
that the analysis we have presented so far is indeed correct.

\begin{table}
\begin{tabular}{|c|c|c|c|c|c|}
\hline 
$P_{0}$&
Eq. \ref{Pressure-mechanical}&
S.E.&
$T$&
Eq. \ref{Temperature-kinetic}&
S.E.\tabularnewline
\hline
\hline 
0.5&
0.50002&
0.0003&
1&
1.0006&
0.001\tabularnewline
\hline 
3.5&
3.5001&
0.0015&
4&
3.9970&
0.003\tabularnewline
\hline
\end{tabular}

\caption{The average values obtained for the pressure and the temperature
along with the standard error for a system with 36 particles in 2
dimensions. \label{table: small equib}}
\end{table}

\section{The Crooks Fluctuation Theorem and the Jarzynski Equality}

The CFT \cite{Crooks-JSP-98,Crooks-PRE-99} gives the probability
of observing an amount of work $\Delta W=B$, in transforming a system
from an initial equilibrium state $I_{E,1}$ to a final state $I_{E,2}$
(where the system will eventually relax to a new equilibrium), relative
to observing the opposite amount of work $\Delta W=-B$ for the reverse
process starting from the equilibrium state $I_{E,2}$. The ratio
between the likelihood of these two observations is given in terms
of the difference in free energy between the two equilibrium states.
The JE \cite{Jarzynski-PRL-97,Jarzynski-PRE-97} gives the change
in free energy between two equilibrium states in terms of the ensemble
average of the exponential of the amount of work it takes to do the
transformation. For both theories the two different equilibrium states
must have the same temperature. These theories stand out as important
due to the fact that they give differences in equilibrium thermodynamic
potentials from nonequilibrium data. They also remain valid arbitrarily
far from equilibrium. 

The Gibbs free energy is given by,\begin{equation}
G=-k_{B}T_{0}\ln\Delta(P_{0},T_{0}),\label{Gibbs Free E 0}\end{equation}
where the partition function $\Delta$ is the normalisation constant
for $f_{0}(\boldsymbol{\Gamma},V)$ (see Eq. \ref{f_0}) and thus,

\begin{equation}
G=-k_{B}T_{0}\ln\left[\int_{0}^{\infty}dV\int d\boldsymbol{\Gamma}\delta(\mathbf{p}_{tot})\delta(\mathbf{R})\exp[-\beta(H_{0}(\boldsymbol{\Gamma})+P_{0}V)]\right].\label{Gibbs Free E}\end{equation}
A natural definition of the rate of work (here the system's temperature
$T_{0}$ is held fixed) done on the system can be obtained by considering
the first law of thermodynamics,\begin{equation}
\frac{dw_{t}}{dt}=\frac{\partial I_{E}(\boldsymbol{\Gamma}^{\prime},t)}{\partial t}=\frac{dI_{E}(\boldsymbol{\Gamma}^{\prime}(t),t)}{dt}-k_{B}T_{0}\Lambda,\label{work}\end{equation}
where the time dependence in the instantaneous enthalpy $I_{E}$ is
due to a parametric change, i.e. $I_{E}(\boldsymbol{\Gamma}^{\prime},t)=I_{E}(\boldsymbol{\Gamma}^{\prime},\lambda(t))$
in the Hamiltonian for the system. This definition remains useful
when the parametric change alters the pressure $P_{0}$. If the time
dependence is solely due to changing the pressure $P_{0}$ then Eq.
\ref{work} gives $\dot{w}_{t}=V\: dP_{0}(t)/dt$. Of course if $P_{0}$
is parametrically changed the blanketing technique is no longer able
to circumvent thermostatting artefacts. At time $t=0$ the system
is in an initial equilibrium with $I_{E}(\boldsymbol{\Gamma}^{\prime},0)=I_{E,1}(\boldsymbol{\Gamma}^{\prime})$,
the system then undergoes some transformation during the time interval
$0<t<\tau$ and then for times $t>\tau$ the system relaxes to a new
equilibrium, $I_{E}(\boldsymbol{\Gamma}^{\prime},t>\tau)=I_{E,2}(\boldsymbol{\Gamma}^{\prime})$,
which is reached at the sufficiently later time $t_{f}\gg\tau$. We
need to calculate the probability of observing a certain amount of
work done on the system in transforming it from an initial equilibrium
with $I_{E,1}$ compared to the opposite amount of work when the system
undergoes the same transformation in reverse from an initial equilibrium
with $I_{E,2}$. We consider an ensemble of trajectories, the number
of which approaches infinity. Following the procedure exploited in
the derivation of the Evans Searles fluctuation theorem \cite{Evans-PRE-94,Evans-adv-phys-02},
the probability of observing a trajectory with selected values of
a property, such as work, can be obtained by summing over the product
of the distribution function and the phase volume $\delta V_{\boldsymbol{\Gamma}}$
for trajectories that satisfy the selection criterion. Hence, the
probability ratio between the forward and reverse process is, \begin{equation}
\frac{p_{f}(w_{\tau}=B)}{p_{r}(w_{\tau}=-B)}=\frac{\sum_{\boldsymbol{\Gamma}(0)|w_{\tau}=B,dB}\delta V_{\boldsymbol{\Gamma}}(\boldsymbol{\Gamma^{\prime}}(0))f_{1}(\boldsymbol{\Gamma^{\prime}}(0))}{\sum_{\boldsymbol{\Gamma}(0)|w_{\tau}=B,dB}\delta V_{\boldsymbol{\Gamma}}(\boldsymbol{\Gamma^{\prime,T}}(\tau))f_{2}(,\boldsymbol{\Gamma}^{\prime T}(\tau))]},\label{p-ratio-0}\end{equation}
where $f_{1}$ is the distribution function for state 1 and $f_{2}$
for state 2, the vector $\boldsymbol{\Gamma}^{\prime,T}$ represents
the time reversal transformation of $\boldsymbol{\Gamma}^{\prime}$
(which has a unit Jacobian) i.e. $\boldsymbol{p}\rightarrow-\boldsymbol{p}$,
$\alpha_{V}\rightarrow-\alpha_{V}$ and $\alpha_{T}\rightarrow-\alpha_{T}$,
and $\delta V(\boldsymbol{\Gamma}^{\prime}(0))$ is the hyper-volume
element in extended phase space surrounding the trajectory bundle.
The summation is over all trajectory bundles which result in the amount
of work done by the complete transformation process being in the range
$w_{\tau}=B\pm dB/2$. It is also assumed here, and below, that all
$\boldsymbol{\Gamma}$ lie on the $\mathbf{p}_{tot}=\mathbf{0}$ and
$\mathbf{q}_{cm}=\mathbf{0}$ hypersurface. For the particular case
here, using the equations of motion Eq. \ref{EOM} this becomes,

\begin{eqnarray}
\frac{p_{f}(w_{\tau}=B)}{p_{r}(w_{\tau}=-B)} & = & \frac{\sum_{\boldsymbol{\Gamma}^{\prime}(0)|w_{\tau}=B,dB}\delta V_{\boldsymbol{\Gamma}}(\boldsymbol{\Gamma}^{\prime}(0))\exp[-\beta I_{E,1}(\boldsymbol{\Gamma}^{\prime}(0))]}{\sum_{\boldsymbol{\Gamma}^{\prime}(0)|w_{\tau}=B,dB}\delta V_{\boldsymbol{\Gamma}}(\boldsymbol{\Gamma}^{\prime,T}(\tau))\exp[-\beta(I_{E,2}(\boldsymbol{\Gamma}^{\prime,T}(\tau))]}\frac{\exp(\beta G_{1})}{\exp(\beta G_{2})},\label{p-ratio-1}\end{eqnarray}
The quantities $G_{1}$ and $G_{2}$ are calculated from Eq. \ref{Gibbs Free E}
with $I_{E}(\boldsymbol{\Gamma}^{\prime})$ given by either $I_{E,1}(\boldsymbol{\Gamma}^{\prime})$
or $I_{E,2}(\boldsymbol{\Gamma}^{\prime})$. The first law of thermodynamics
manifests itself in Eq. \ref{work} and thus the rate at which heat
is exchanged with the reservoir is given by $k_{B}T\Lambda$. The
probability distribution, in the streaming representation, starting
from an initial distribution at time $t=0$ may be obtained from the
streaming Liouville equation \cite{Evans-PRE-95,Evans-adv-phys-02}
and is\begin{equation}
f(\boldsymbol{\Gamma}^{\prime}(t),t)=f(\boldsymbol{\Gamma}^{\prime}(0),0)\exp\left(-\int_{0}^{t}ds\,\Lambda(\boldsymbol{\Gamma}^{\prime}(s))\right).\label{streaming-dist}\end{equation}
If we consider a packet of trajectories initially contained by the
infinitesimal element of volume $\delta V(\boldsymbol{\Gamma}^{\prime}(0))$
we may later observe them contained by the infinitesimal element of
volume $\delta V(\boldsymbol{\Gamma}^{\prime}(t))$ . By construction,
the number of trajectories, or ensemble members, in these two infinitesimal
elements is conserved and thus we have\begin{equation}
\delta V(\boldsymbol{\Gamma}^{\prime}(t))f(\boldsymbol{\Gamma}^{\prime}(t))=\delta V(\boldsymbol{\Gamma}^{\prime}(0))f(\boldsymbol{\Gamma}^{\prime}(0))\label{phase-volume-1}\end{equation}
and in turn\begin{equation}
\delta V(\boldsymbol{\Gamma}^{\prime}(t))=\delta V(\boldsymbol{\Gamma}^{\prime}(0))\exp\left(\int_{0}^{t}ds\,\Lambda(\boldsymbol{\Gamma}^{\prime}(s))\right)\label{phase-volume-2}\end{equation}
Now by the first law Eq. \ref{work} we have $I_{E}(\tau)=I_{E}(0)+w_{\tau}+k_{B}T\int_{0}^{t}ds\,\Lambda(\boldsymbol{\Gamma}^{\prime}(s))$:
substituting this into the denominator of Eq. \ref{p-ratio-1} along
with Eq. \ref{phase-volume-2} and noting $I_{E}(\boldsymbol{\Gamma}^{\prime,T})=I_{E}(\boldsymbol{\Gamma}^{\prime})$
gives,\begin{equation}
\frac{p_{f}(w_{\tau}=B)}{p_{r}(w_{\tau}=-B)}=\frac{\sum_{\boldsymbol{\Gamma}^{\prime}(0)|w_{\tau}=B,dB}\delta V(\boldsymbol{\Gamma}^{\prime}(0))\exp[-\beta I_{E,1}(\boldsymbol{\Gamma}^{\prime}(0))]}{\sum_{\boldsymbol{\Gamma}^{\prime}(0)|w_{\tau}=B,dB}\delta V(\boldsymbol{\Gamma}^{\prime}(0))\exp[-\beta(I_{E,1}(\boldsymbol{\Gamma}^{\prime}(0))+w_{\tau})]}\exp(-\beta\Delta G)\label{p-ratio-2}\end{equation}
where $\Delta G=G_{2}-G_{1}$. Thus we obtain the CFT in the isothermal
isobaric ensemble,\begin{equation}
\frac{p_{f}(w_{\tau}=B)}{p_{r}(w_{\tau}=-B)}=\exp[-\beta(\Delta G-B)]\label{Crooks}\end{equation}
which gives the probability of observing the amount of work $w_{\tau}=B$
done in the transformation process from initial equilibrium state
$I_{E,1}$ relative to the probability of observing the amount of
work $w_{\tau}=-B$ for the reverse process starting from an initial
equilibrium state $I_{E,2}$. It is now trivial to integrate Eq. \ref{Crooks},\begin{equation}
\int_{-\infty}^{\infty}dB\, p_{f}(w_{\tau}=B)\exp[-\beta B]=\exp[-\beta\Delta G]\int_{-\infty}^{\infty}dB\, p_{r}(w_{\tau}=-B)\label{Crooks-JE}\end{equation}
and arrive at the JE\begin{equation}
\left\langle \exp[-\beta w_{\tau}]\right\rangle =\exp[-\beta\Delta G].\label{JE}\end{equation}
Thus by taking the ensemble average of the exponential of the work
done (defined through Eq. \ref{work}) in transforming the system
we obtain the change in Gibbs free energy $\Delta G$. We again point
out that this derivation, which starts from the basic equations of
motion, has been carried out for a system of interest governed by
the fundamental Newtonian equations of motion. This is only possible
because the equations of motion we have introduced here allow us to
in effect decouple the Newtonian system of interest from the larger
thermal reservoir and take advantage of the principle of locality.
This proof is valid far from equilibrium as opposed to existing stochastic
proofs for the isothermal isobaric ensemble \cite{Park-JCP-04} which
are only valid for Markovian systems.

The JE and CFT provide practical approaches for obtaining $\Delta G$
by measuring the work done along an ensemble of nonequilibrium paths
that transform the system from one state to another. However since
the derivation of Eq. \ref{JE} is reliant on the specification of
the equilibrium distribution function, numerical verification of this
equation is also an indication that the distribution function indicated
in Eq. \ref{equib-dist-function} is actually generated by the equilibrium
equations of motion.

\section{Demonstration: Stretching a Crude Polymer Model}

\begin{figure}
\resizebox{8.5cm}{!}{\includegraphics{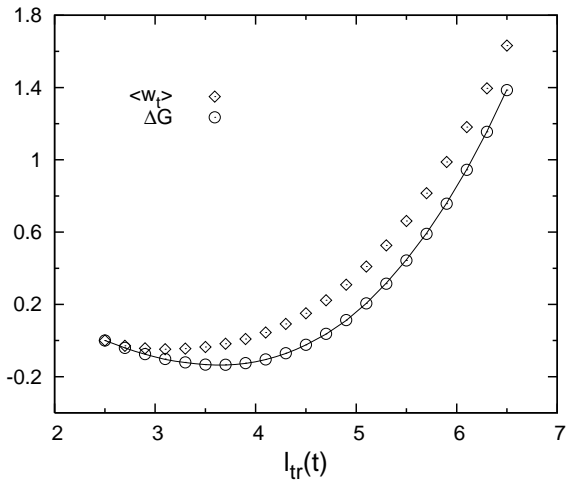}}

\caption{The solid line is the change in Gibbs free energy as a function of
distance between the two traps $l_{tr}$ calculated from a series
of equilibrium simulations using Eq. \ref{dG_dlt}. The diamonds are
the work done in stretching the polymer at a finite rate and the circles
are the results obtained from the JE, Eq. \ref{JE}.\label{figure-JE}}
\end{figure}
 To form a crude model of a single polymer suspended in a solvent
of Cartesian dimension $D=2$ we use the finite extendable nonlinear
elastic (FENE) chain model \cite{kroger-pr-04}. The two end particles
of the polymer are held by a pair of harmonic wells $U_{h}(x)=\frac{1}{2}k_{t}x^{2}$,
i.e. as in optical traps \cite{wang-prl-02}. We may then apply the
JE to the case where the polymer is stretched due to the two traps
being moved apart. We note that previous work has addressed how to
determine the free energy difference as a function of the polymer
separation rather than the trap separation \textbf{\cite{Hummer-acr-05}}.
All particles, solvent and polymer, interact by the same repulsive
WCA potential \cite{WCA-JCP-71} and have the same mass. In addition
neighbouring polymer particles are bound together by the FENE potential
\cite{kroger-pr-04} \begin{equation}
U_{FENE}=-\frac{1}{2}Kr_{0}^{2}\ln\left[1-(\Delta r/r_{0})^{2}\right],\label{U_FENE}\end{equation}
where the particles must be separated by a distance $\Delta r$ which
is less than $r_{0}$ where the potential diverges and $K$ is the
strength of the potential. The spatial derivative gives us the force\begin{equation}
F_{FENE}=-\nabla U_{FENE}=-K\Delta\mathbf{r}/\left(1-(\Delta r/r_{0})^{2}\right).\label{F_FENE}\end{equation}
We set the FENE parameters to $K=10\epsilon$ and $r_{0}=1.5\sigma$
where $\epsilon$ and $\sigma$ are the parameters from the WCA potential.
The total number of particles is $N=300$, the number of polymer particles
is $N_{p}=8$, the trap strength is $k_{t}=2\epsilon/\sigma^{2}$,
the temperature is $T=\epsilon/k_{B}$ were $k_{B}$ is Boltzmann's
constant and the pressure is $P_{0}=2\sigma^{2}/\epsilon$. The time
is reported in units of $\sigma\sqrt{m/\epsilon}$ where $m$ is the
particle mass and lengths are reported in units of $\sigma$. The
optical traps where initially set at a separation distance of $l_{tr}=2.5$
and stretched to a distance of $l_{tr}=6.5$. With $l_{tr}=2.5$ the
average volume was found to be $\left\langle V\right\rangle =521.9\pm0.2$
and with $l_{tr}=6.5$, $\left\langle V\right\rangle =522.3\pm0.2$,
where the figure in the error estimate is one standard error. The
traps were centred abound the Newtonian region in the simulation.
In the spirit of the method originally introduced by Hoover and Ree
\cite{HOOVER-JCP-68} the average equilibrium length of the polymer
$l_{p}$, in the direction of the vector connecting the two traps,
was computed for $31$ separation distances between and including
the shortest and longest length. This data was then used to compute
the equilibrium change in Gibbs free energy using the equation\begin{equation}
\frac{dG}{dl_{tr}}=-\frac{1}{2}k_{tr}\left(l_{p}-l_{tr}\right),\label{dG_dlt}\end{equation}
with the trapezoidal quadrature. For the JE the work for Eq. \ref{JE}
obtained from Eq. \ref{work} is computed from the equation\begin{equation}
\frac{dw_{t}}{dt}=-\frac{1}{2}k_{tr}\frac{dl_{t}}{dt}\left(x_{p,r}-x_{p,l}-l_{tr}\right),\label{work-JE-polymer}\end{equation}
where $x_{p,r}$ is the $x$ component of the position of the particle
at the right hand end of the polymer and $x_{p,l}$ that at the left
hand end. Both traps are located on the $x$ axis. 

A total of $5\times10^{4}$ nonequilibrium trajectories were computed
starting from equilibrium with $l_{tr}=2.5$ and stretched at a constant
rate until $l_{tr}=6.5$ over a duration of $\tau=100$. The results
may be seen in Fig. \ref{figure-JE}. The average work done in the
nonequilibrium stretching of the polymer is clearly greater than the
change in free energy indicating that energy is being irreversibly
dissipated into the thermal reservoir. It can be seen, despite the
irreversible work, that there is excellent agreement between the change
in free energy calculated using equilibrium methods Eq. \ref{dG_dlt}
and the change in free energy calculated using the JE Eqs. \ref{JE}
\& \ref{work-JE-polymer}.

\section{Conclusions}

We have introduced equations of motion that preserve the isothermal
isobaric ensemble's distribution function yet still feature a region
which is governed by the natural Newtonian equations of motion. This
natural region can be chosen to be arbitrarily large. As we have argued
previously \cite{Evans-adv-phys-02,Williams-PRE-04} if the thermostatted,
barostatted blanket is far enough removed (from the system of interest)
its details can not possibly effect the physical observations made
in the system of interest. This amounts to the assumption of locality,
which is one of the most important and well established assumptions
in physics. When a far from equilibrium process occurs in the system
of interest the heat flux which is dissipated into the outer blanket
diminishes with distance: for $D=3$ this will be proportional to
the inverse square of the distance. If the blanket is far enough removed
from the system of interest the thermostat and barostat will only
act on a region that is in local equilibrium and it is known that
this class of thermostatted dynamics does not introduce any artifacts
when in local equilibrium \cite{Evans-Morriss-book}. 

In the arrangement described here, the system of interest is in a
spherical region. Other physical arrangements could be designed, for
example it could be a slit shaped region. This arrangement would be
more useful for studying some nonequilibrium steady state dynamics
(e.g. Poiseuille flow). 

Using these equations of motion we have derived the CFT and the JE
in the isothermal isobaric ensemble that will be applicable regardless
of how far from equilibrium the system of interest is driven. The
only physical assumption necessary for this is the assumption of locality.
Existing derivations for the isothermal isobaric ensemble using Markovian
stochastic dynamics \cite{Park-JCP-04}, can only be linked to the
fundamental microscopic equations, at best, in the near equilibrium
linear response regime. One of the remarkable features of these theories
is that they remain valid far from equilibrium. We now have a proofs
of the JE and the CFT that remain valid far from equilibrium in the
isothermal isobaric ensemble. It is this ensemble, with the Gibbs
free energy as the thermodynamic potential, which is most relevant
to physical experiments and processes.

\begin{acknowledgments}
We thank the Australian Research Council for financial support and
the Australian Partnership for Advanced Computing for computational
facilities. We also thank Emil Mittag and Peter Daivis for helpful
discussions.
\end{acknowledgments}
\bibliographystyle{apsrev}
\bibliography{conpres}

\begin{thebibliography}{35}
\expandafter\ifx\csname natexlab\endcsname\relax\def\natexlab#1{#1}\fi
\expandafter\ifx\csname bibnamefont\endcsname\relax
  \def\bibnamefont#1{#1}\fi
\expandafter\ifx\csname bibfnamefont\endcsname\relax
  \def\bibfnamefont#1{#1}\fi
\expandafter\ifx\csname citenamefont\endcsname\relax
  \def\citenamefont#1{#1}\fi
\expandafter\ifx\csname url\endcsname\relax
  \def\url#1{\texttt{#1}}\fi
\expandafter\ifx\csname urlprefix\endcsname\relax\def\urlprefix{URL }\fi
\providecommand{\bibinfo}[2]{#2}
\providecommand{\eprint}[2][]{\url{#2}}

\bibitem[{\citenamefont{Evans et~al.}(1993)\citenamefont{Evans, Cohen, and
  Morriss}}]{Evans-PRL-93}
\bibinfo{author}{\bibfnamefont{D.~J.} \bibnamefont{Evans}},
  \bibinfo{author}{\bibfnamefont{E.~G.~D.} \bibnamefont{Cohen}},
  \bibnamefont{and} \bibinfo{author}{\bibfnamefont{G.~P.}
  \bibnamefont{Morriss}}, \bibinfo{journal}{Phys. Rev. Lett.}
  \textbf{\bibinfo{volume}{71}}, \bibinfo{pages}{2401} (\bibinfo{year}{1993}).

\bibitem[{\citenamefont{Evans and Searles}(1994)}]{Evans-PRE-94}
\bibinfo{author}{\bibfnamefont{D.~J.} \bibnamefont{Evans}} \bibnamefont{and}
  \bibinfo{author}{\bibfnamefont{D.~J.} \bibnamefont{Searles}},
  \bibinfo{journal}{Phys. Rev. E} \textbf{\bibinfo{volume}{50}},
  \bibinfo{pages}{1645} (\bibinfo{year}{1994}).

\bibitem[{\citenamefont{Gallavotti and Cohen}(1995)}]{Gallavotti-PRL-95}
\bibinfo{author}{\bibfnamefont{G.}~\bibnamefont{Gallavotti}} \bibnamefont{and}
  \bibinfo{author}{\bibfnamefont{E.~G.~D.} \bibnamefont{Cohen}},
  \bibinfo{journal}{Phys. Rev. Lett.} \textbf{\bibinfo{volume}{74}},
  \bibinfo{pages}{2694} (\bibinfo{year}{1995}).

\bibitem[{\citenamefont{Jarzynski}(1997{\natexlab{a}})}]{Jarzynski-PRL-97}
\bibinfo{author}{\bibfnamefont{C.}~\bibnamefont{Jarzynski}},
  \bibinfo{journal}{Phys. Rev. Lett.} \textbf{\bibinfo{volume}{78}},
  \bibinfo{pages}{2690} (\bibinfo{year}{1997}{\natexlab{a}}).

\bibitem[{\citenamefont{Crooks}(1998)}]{Crooks-JSP-98}
\bibinfo{author}{\bibfnamefont{G.~E.} \bibnamefont{Crooks}},
  \bibinfo{journal}{J. Stat. Phys.} \textbf{\bibinfo{volume}{90}},
  \bibinfo{pages}{1481} (\bibinfo{year}{1998}).

\bibitem[{\citenamefont{Park and Schulten}(2004)}]{Park-JCP-04}
\bibinfo{author}{\bibfnamefont{S.}~\bibnamefont{Park}} \bibnamefont{and}
  \bibinfo{author}{\bibfnamefont{K.}~\bibnamefont{Schulten}},
  \bibinfo{journal}{J. Chem. Phys.} \textbf{\bibinfo{volume}{120}},
  \bibinfo{pages}{5946} (\bibinfo{year}{2004}).

\bibitem[{\citenamefont{Cuendet}(2006{\natexlab{a}})}]{Cuendet-JCP-06}
\bibinfo{author}{\bibfnamefont{M.~A.} \bibnamefont{Cuendet}},
  \bibinfo{journal}{J. Chem. Phys.} \textbf{\bibinfo{volume}{125}},
  \bibinfo{pages}{144109} (\bibinfo{year}{2006}{\natexlab{a}}).

\bibitem[{\citenamefont{Evans}(2003)}]{Evans-molphys-03}
\bibinfo{author}{\bibfnamefont{D.~J.} \bibnamefont{Evans}},
  \bibinfo{journal}{Mol. Phys.} \textbf{\bibinfo{volume}{101}},
  \bibinfo{pages}{1551} (\bibinfo{year}{2003}).

\bibitem[{\citenamefont{Cuendet}(2006{\natexlab{b}})}]{Cuendet-PRL-06}
\bibinfo{author}{\bibfnamefont{M.~A.} \bibnamefont{Cuendet}},
  \bibinfo{journal}{Phys. Rev. Lett.} \textbf{\bibinfo{volume}{96}},
  \bibinfo{pages}{120602} (\bibinfo{year}{2006}{\natexlab{b}}).

\bibitem[{\citenamefont{Sch\"oll-Paschinger and Dellago}(2006)}]{scholl-JCP-06}
\bibinfo{author}{\bibfnamefont{E.}~\bibnamefont{Sch\"oll-Paschinger}}
  \bibnamefont{and} \bibinfo{author}{\bibfnamefont{C.}~\bibnamefont{Dellago}},
  \bibinfo{journal}{J. Chem. Phys.} \textbf{\bibinfo{volume}{125}},
  \bibinfo{pages}{054105} (\bibinfo{year}{2006}).

\bibitem[{\citenamefont{Jarzynski}(1997{\natexlab{b}})}]{Jarzynski-PRE-97}
\bibinfo{author}{\bibfnamefont{C.}~\bibnamefont{Jarzynski}},
  \bibinfo{journal}{Phys. Rev. E} \textbf{\bibinfo{volume}{56}},
  \bibinfo{pages}{5018} (\bibinfo{year}{1997}{\natexlab{b}}).

\bibitem[{\citenamefont{Liphardt et~al.}(2002)\citenamefont{Liphardt, Dumont,
  Smith, Tinoco, and Bustamante}}]{Liphardt-sci-02}
\bibinfo{author}{\bibfnamefont{J.}~\bibnamefont{Liphardt}},
  \bibinfo{author}{\bibfnamefont{S.}~\bibnamefont{Dumont}},
  \bibinfo{author}{\bibfnamefont{S.~B.} \bibnamefont{Smith}},
  \bibinfo{author}{\bibfnamefont{I.}~\bibnamefont{Tinoco}}, \bibnamefont{and}
  \bibinfo{author}{\bibfnamefont{C.}~\bibnamefont{Bustamante}},
  \bibinfo{journal}{Science} \textbf{\bibinfo{volume}{296}},
  \bibinfo{pages}{1832} (\bibinfo{year}{2002}).

\bibitem[{\citenamefont{Collin et~al.}(2005)\citenamefont{Collin, Ritort,
  Jarzynski, Smith, Tinoco, and Bustamante}}]{Collin-Nat-05}
\bibinfo{author}{\bibfnamefont{D.}~\bibnamefont{Collin}},
  \bibinfo{author}{\bibfnamefont{F.}~\bibnamefont{Ritort}},
  \bibinfo{author}{\bibfnamefont{C.}~\bibnamefont{Jarzynski}},
  \bibinfo{author}{\bibfnamefont{S.~B.} \bibnamefont{Smith}},
  \bibinfo{author}{\bibfnamefont{I.}~\bibnamefont{Tinoco}}, \bibnamefont{and}
  \bibinfo{author}{\bibfnamefont{C.}~\bibnamefont{Bustamante}},
  \bibinfo{journal}{Nature} \textbf{\bibinfo{volume}{437}},
  \bibinfo{pages}{231} (\bibinfo{year}{2005}).

\bibitem[{\citenamefont{Hummer and Szabo}(2001)}]{Hummer-pnas-01}
\bibinfo{author}{\bibfnamefont{G.}~\bibnamefont{Hummer}} \bibnamefont{and}
  \bibinfo{author}{\bibfnamefont{A.}~\bibnamefont{Szabo}},
  \bibinfo{journal}{PNAS} \textbf{\bibinfo{volume}{98}}, \bibinfo{pages}{3658}
  (\bibinfo{year}{2001}).

\bibitem[{\citenamefont{Zwanzig}(2001)}]{zwanzig-book}
\bibinfo{author}{\bibfnamefont{R.}~\bibnamefont{Zwanzig}},
  \emph{\bibinfo{title}{Nonequilibrium Statistical Mechanics}}
  (\bibinfo{publisher}{Oxford University Press}, \bibinfo{address}{Oxford},
  \bibinfo{year}{2001}).

\bibitem[{\citenamefont{Evans and Morriss}(1990)}]{Evans-Morriss-book}
\bibinfo{author}{\bibfnamefont{D.~J.} \bibnamefont{Evans}} \bibnamefont{and}
  \bibinfo{author}{\bibfnamefont{G.~P.} \bibnamefont{Morriss}},
  \emph{\bibinfo{title}{Statistical Mechanics of Nonequilibrium Liquids.}}
  (\bibinfo{publisher}{Academic}, \bibinfo{address}{London},
  \bibinfo{year}{1990}), \bibinfo{note}{also available at:
  http://rsc.anu.edu.au/$\sim$evans/evansmorrissbook.htm}.

\bibitem[{\citenamefont{Hansen and McDonald}(1996)}]{Hansen-book}
\bibinfo{author}{\bibfnamefont{J.~P.} \bibnamefont{Hansen}} \bibnamefont{and}
  \bibinfo{author}{\bibfnamefont{I.~R.} \bibnamefont{McDonald}},
  \emph{\bibinfo{title}{Theory of Simple Liquids}}
  (\bibinfo{publisher}{Academic Press}, \bibinfo{year}{1996}),
  \bibinfo{edition}{2nd} ed., \bibinfo{note}{see especially p309-310}.

\bibitem[{\citenamefont{Hummer and Szabo}(2005)}]{Hummer-acr-05}
\bibinfo{author}{\bibfnamefont{G.}~\bibnamefont{Hummer}} \bibnamefont{and}
  \bibinfo{author}{\bibfnamefont{A.}~\bibnamefont{Szabo}},
  \bibinfo{journal}{Acc. Chem. Res.} \textbf{\bibinfo{volume}{38}},
  \bibinfo{pages}{504} (\bibinfo{year}{2005}).

\bibitem[{\citenamefont{Crooks}(1999)}]{Crooks-PRE-99}
\bibinfo{author}{\bibfnamefont{G.~E.} \bibnamefont{Crooks}},
  \bibinfo{journal}{Phys. Rev. E} \textbf{\bibinfo{volume}{60}},
  \bibinfo{pages}{2721} (\bibinfo{year}{1999}).

\bibitem[{\citenamefont{Williams et~al.}(2004)\citenamefont{Williams, Searles,
  and Evans}}]{Williams-PRE-04}
\bibinfo{author}{\bibfnamefont{S.~R.} \bibnamefont{Williams}},
  \bibinfo{author}{\bibfnamefont{D.~J.} \bibnamefont{Searles}},
  \bibnamefont{and} \bibinfo{author}{\bibfnamefont{D.~J.} \bibnamefont{Evans}},
  \bibinfo{journal}{Phys. Rev. E} \textbf{\bibinfo{volume}{70}},
  \bibinfo{pages}{066113} (\bibinfo{year}{2004}).

\bibitem[{\citenamefont{Ayton et~al.}(2001)\citenamefont{Ayton, Evans, and
  Searles}}]{Ayton-JCP-01}
\bibinfo{author}{\bibfnamefont{G.}~\bibnamefont{Ayton}},
  \bibinfo{author}{\bibfnamefont{D.~J.} \bibnamefont{Evans}}, \bibnamefont{and}
  \bibinfo{author}{\bibfnamefont{D.~J.} \bibnamefont{Searles}},
  \bibinfo{journal}{J. Chem. Phys.} \textbf{\bibinfo{volume}{115}},
  \bibinfo{pages}{2033} (\bibinfo{year}{2001}).

\bibitem[{\citenamefont{Evans and Searles}(2002)}]{Evans-adv-phys-02}
\bibinfo{author}{\bibfnamefont{D.~J.} \bibnamefont{Evans}} \bibnamefont{and}
  \bibinfo{author}{\bibfnamefont{D.~J.} \bibnamefont{Searles}},
  \bibinfo{journal}{Adv. Phys.} \textbf{\bibinfo{volume}{51}},
  \bibinfo{pages}{1529} (\bibinfo{year}{2002}).

\bibitem[{\citenamefont{Nos\'e}(1984{\natexlab{a}})}]{Nose-JCP-84}
\bibinfo{author}{\bibfnamefont{S.}~\bibnamefont{Nos\'e}}, \bibinfo{journal}{J.
  Chem. Phys.} \textbf{\bibinfo{volume}{81}}, \bibinfo{pages}{511}
  (\bibinfo{year}{1984}{\natexlab{a}}).

\bibitem[{\citenamefont{Nos\'e}(1984{\natexlab{b}})}]{Nose-mp-84}
\bibinfo{author}{\bibfnamefont{S.}~\bibnamefont{Nos\'e}},
  \bibinfo{journal}{Mol. Phys.} \textbf{\bibinfo{volume}{52}},
  \bibinfo{pages}{255} (\bibinfo{year}{1984}{\natexlab{b}}).

\bibitem[{\citenamefont{Hoover}(1985)}]{Hoover-PRA-85}
\bibinfo{author}{\bibfnamefont{W.~G.} \bibnamefont{Hoover}},
  \bibinfo{journal}{Phys. Rev. A.} \textbf{\bibinfo{volume}{31}},
  \bibinfo{pages}{1695} (\bibinfo{year}{1985}).

\bibitem[{foo({\natexlab{a}})}]{foot1}
\bibinfo{note}{If the centre of the simulation cell is not at the origin, a
  simple translation of the coordinates can be used to relate the positions to
  the coordinate system used here. To properly determine the function
  $g(\mathbf{q}_{i},V(t))$ we must consider the scalar distance $q_i$ from the
  centre of the current unit cell}.

\bibitem[{\citenamefont{Rapaport}(2004)}]{rapaport-book}
\bibinfo{author}{\bibfnamefont{D.~C.} \bibnamefont{Rapaport}},
  \emph{\bibinfo{title}{The Art of Molecular Dynamics Simulation}}
  (\bibinfo{publisher}{Cambridge University Press}, \bibinfo{year}{2004}).

\bibitem[{\citenamefont{Braga and Travis}(2006)}]{Braga-JCP-2006}
\bibinfo{author}{\bibfnamefont{C.}~\bibnamefont{Braga}} \bibnamefont{and}
  \bibinfo{author}{\bibfnamefont{K.~P.} \bibnamefont{Travis}},
  \bibinfo{journal}{J. Chem. Phys.} \textbf{\bibinfo{volume}{124}},
  \bibinfo{pages}{104102} (\bibinfo{year}{2006}).

\bibitem[{\citenamefont{Wang et~al.}(2002)\citenamefont{Wang, Sevick, Mittag,
  Searles, and Evans}}]{wang-prl-02}
\bibinfo{author}{\bibfnamefont{G.~M.} \bibnamefont{Wang}},
  \bibinfo{author}{\bibfnamefont{E.~M.} \bibnamefont{Sevick}},
  \bibinfo{author}{\bibfnamefont{E.}~\bibnamefont{Mittag}},
  \bibinfo{author}{\bibfnamefont{D.~J.} \bibnamefont{Searles}},
  \bibnamefont{and} \bibinfo{author}{\bibfnamefont{D.~J.} \bibnamefont{Evans}},
  \bibinfo{journal}{Phys. Rev. Lett.} \textbf{\bibinfo{volume}{89}},
  \bibinfo{pages}{050601} (\bibinfo{year}{2002}).

\bibitem[{foo({\natexlab{b}})}]{foot2}
\bibinfo{note}{For a periodic system the sum $\sum_i^N
  g(\boldsymbol{q}_i,V)\boldsymbol{F}_i\cdot\boldsymbol{q}_i$ must be replaced
  with an expression that respects the minimum image condition. For pairwise
  additive potentials this is $\sum_{i=1}^{N-1}\sum_{j=i+1}^{N}
  \boldsymbol{F}_{ij}\cdot(\mathbf{q}_{i}g(\mathbf{q}_{i},V)-(\mathbf{q}_{i}+
  \mathbf{q}_{ij})g(\mathbf{q}_{j},V))$ where $\mathbf{F}_{ij}$ is the force
  and $\mathbf{q}_{ij}$ is the displacement between a pair of minimum image
  particles and $\mathbf{q}_i$ and $\mathbf{q}_j$ are in the central cell. For
  particles which interact across a boundary the function $g$ is such that
  $g(\mathbf{q}_{i},V)=g(\mathbf{q}_{j},V)=1$.}

\bibitem[{\citenamefont{Weeks et~al.}(1971)\citenamefont{Weeks, Chandler, and
  Andersen}}]{WCA-JCP-71}
\bibinfo{author}{\bibfnamefont{J.~D.} \bibnamefont{Weeks}},
  \bibinfo{author}{\bibfnamefont{D.}~\bibnamefont{Chandler}}, \bibnamefont{and}
  \bibinfo{author}{\bibfnamefont{H.~C.} \bibnamefont{Andersen}},
  \bibinfo{journal}{J. Chem. Phys.} \textbf{\bibinfo{volume}{54}},
  \bibinfo{pages}{5237} (\bibinfo{year}{1971}).

\bibitem[{\citenamefont{Butcher and Wanner}(1996)}]{butcher-ANM-96}
\bibinfo{author}{\bibfnamefont{J.~C.} \bibnamefont{Butcher}} \bibnamefont{and}
  \bibinfo{author}{\bibfnamefont{G.}~\bibnamefont{Wanner}},
  \bibinfo{journal}{Appl. Numer. Math.} \textbf{\bibinfo{volume}{22}},
  \bibinfo{pages}{113} (\bibinfo{year}{1996}).

\bibitem[{\citenamefont{Evans and Searles}(1995)}]{Evans-PRE-95}
\bibinfo{author}{\bibfnamefont{D.~J.} \bibnamefont{Evans}} \bibnamefont{and}
  \bibinfo{author}{\bibfnamefont{D.~J.} \bibnamefont{Searles}},
  \bibinfo{journal}{Phys. Rev. E} \textbf{\bibinfo{volume}{52}},
  \bibinfo{pages}{5839} (\bibinfo{year}{1995}).

\bibitem[{\citenamefont{Kr\"oger}(2004)}]{kroger-pr-04}
\bibinfo{author}{\bibfnamefont{M.}~\bibnamefont{Kr\"oger}},
  \bibinfo{journal}{Phys. Rep.} \textbf{\bibinfo{volume}{390}},
  \bibinfo{pages}{453} (\bibinfo{year}{2004}).

\bibitem[{\citenamefont{Hoover and Ree}(1968)}]{HOOVER-JCP-68}
\bibinfo{author}{\bibfnamefont{W.~G.} \bibnamefont{Hoover}} \bibnamefont{and}
  \bibinfo{author}{\bibfnamefont{F.~H.} \bibnamefont{Ree}},
  \bibinfo{journal}{J. Chem. Phys.} \textbf{\bibinfo{volume}{49}},
  \bibinfo{pages}{3609} (\bibinfo{year}{1968}).

\end{thebibliography}

\end{document}